\documentclass[journal]{IEEEtran}
\ifCLASSINFOpdf
  \usepackage[pdftex]{graphicx}
\else
\fi
\usepackage{multirow}

\usepackage{dblfloatfix}

\usepackage{ulem}

\usepackage{color}

\begin{document}
%
\title{MH370 Burst Frequency Offset Analysis and Implications on Descent Rate at End-of-Flight}
%
%
%

\author{Ian~D.~Holland\\Defence Science and Technology Group, Edinburgh, Australia 
\thanks{Copyright \textcopyright~2018 Commonwealth of Australia.}}

%
%

\markboth{}%
{Holland: MH370 Burst Frequency Offset Analysis and Implications on Descent Rate at End-of-Flight}
%



\maketitle



%
\IEEEpeerreviewmaketitle

\section{Introduction}
%
%
%
%
\IEEEPARstart{O}{n} 7 March 2014 at 16:41:43Z\footnote{All times given in this
article are in UTC or ``Zulu'' time, denoted by ``Z''.}, Malaysian Airlines 
flight MH370 
departed Kuala Lumpur (KL) International Airport bound for Beijing. Less than 
an hour later, following the last recorded radio transmission from MH370 at 
17:19:30Z, the plane's secondary radar transponder went offline. As evidenced by
Malaysian military radar, the plane (registration number 9M-MRO) then veered 
off course unexpectedly, back-tracked across the Malaysian Peninsula, and was 
then tracked heading northwest from Penang through the Malacca Straits. After 
disappearing from radar at 18:22:12Z, it re-established a satellite
communications (SATCOM) link with the Inmarsat satellite I-3F1 at 18:25:27Z.
By analyzing a series of automated messages exchanged via that satellite 
between 
the plane and an Inmarsat ground station in Perth, Australia, it was
determined that the plane continued to fly for six hours, 
before finally ceasing message exchange with the ground station at 00:19:37Z on 
8 March 2014. This article discusses specifically the analysis of
burst frequency offset (BFO) metadata from the SATCOM messages. Importantly,
it is shown that the BFOs corresponding to the last two SATCOM messages from the plane at 00:19:29Z and 00:19:37Z suggest that flight MH370 was rapidly descending and accelerating downwards when message exchange with the ground
station ceased.

An initial analysis by Inmarsat of the SATCOM metadata for MH370 in the last six hours of flight suggested that MH370 had 
flown into the Southern Indian Ocean before SATCOM was ultimately lost (see \cite{Ashton2014} for further details). As
summarized in \cite{Davey2016} an intensive
aerial and surface search was undertaken in the Southern Indian Ocean by an
international search team during March and April 2014, with no MH370 related 
debris found. On 28 April 2014, the aerial search concluded and the search 
transitioned to an underwater phase \cite{Davey2016}. The Australian Transport 
Safety Bureau (ATSB) took responsibility for the definition of the underwater 
search zone. It convened an international flight path prediction working group 
bringing together experts in satellite communications, statistical data 
processing and aviation, in order to estimate the most likely final location 
of flight MH370. The group consisted of representatives from the Australian 
Defence Science and Technology (DST) Group and the other organizations listed 
in the Acknowledgment section of this article.

New methods of analyzing the Inmarsat data were developed by the group, 
resulting 
in the release of reports concerning the likely final location of flight MH370 from 
the ATSB in August 2014 \cite{ATSB2014}, 
October 2014 \cite{ATSB2014a}, and December 2015 \cite{ATSB2015}. 
Inmarsat also published an article regarding their contribution to the flight 
path reconstruction effort \cite{Ashton2014}. The DST Group contribution
that assisted in the definition of an extended 
priority search area in December 2015 \cite{ATSB2015} has been detailed 
in \cite{Davey2016}. This demonstrated how Bayesian analysis was used to
identify a high probability region of where the plane was believed to be
at the time of last SATCOM transmission (00:19:37Z 8 March 2014). The DST Group 
Bayesian method used a prior probability distribution defined by the Malaysian 
military radar, a likelihood function 
describing the relationship between SATCOM measurements and the aircraft 
position and velocity during the flight, and a model of the aircraft dynamics.

It should be noted that this article does not
cover the Bayesian method used in defining the underwater search area. For details on that method, the reader is referred to \cite{Davey2016}.
Instead, this article focuses on:
\begin{enumerate}
	\item A brief review of the statistical analysis of BFOs for several 
	previous flights of 9M-MRO.
	\item Examination of the effects of the plane's vertical velocity
	on the BFO.
	\item An analysis of the effect of the track angle of the plane
	on the BFO towards the end-of-flight.
	\item The establishment of a BFO trend throughout the
	last 6 hours of flight so as to determine an expected BFO at end-of-flight.
	\item An analysis of the behavior of the frequency oscillator in the 
	plane's 
	satellite data unit (SDU) after power outage events such as those 
	believed to have occurred twice during flight MH370, and how this affects 
	the BFO.
	\item An analysis that shows that the final two BFO's are consistent with
	MH370 being in a rapid descent and accelerating downwards.	
\end{enumerate}

The remainder of this article is structured as follows. 
Section~\ref{sec:Timeline} presents a brief timeline of key events 
during the MH370 accident flight. In Section~\ref{sec:SATCOM_Model}, a 
review of the SATCOM model is provided, along
with a brief review of the BFO statistics. This serves as a summary
of work previously presented in \cite{Davey2016}.
Section~\ref{sec:BFO_Detail} describes the effects of aircraft position
and velocity on the BFO.
The settling behavior of the SDU's frequency oscillator for 9M-MRO after 
power-up is then detailed in Section~\ref{sec:OCXO_Detail}. This is important
because the SDU is believed to have undergone a power outage between 00:11Z 
and 00:19Z on 8 Mar 2014, immediately preceding the last two SATCOM 
transmissions from MH370. The effect of oscillator warm-up on 
the BFO after power-up can be used in bounding the descent
rates for MH370 at 00:19:29Z and 00:19:37Z. In Section~\ref{sec:Descent} an 
analysis of the descent rate of MH370 based on the last two BFO values is
presented. This analysis derives lower and upper bounds on the descent 
rate at 00:19:29Z and 00:19:37Z. Conclusions are presented in 
Section~\ref{sec:Conclusions}.
\section{Timeline of Events During the Accident Flight}\label{sec:Timeline}
A comprehensive description of the events occurring during the MH370 accident
flight is provided in \cite{Malaysia2015}.
A summary of salient events to this article is provided
in Table~\ref{tab:Timeline}.

\begin{table*}[!t]
\renewcommand{\arraystretch}{1.3}
\caption{Timeline of Events During MH370}
\label{tab:Timeline}
\centering
\begin{tabular}{|c||c|c|}
\hline
Date and Timestamp(s) & Description & Comments\\
\hline\hline
7 Mar. 2014 16:42Z &
MH370 departs Kuala Lumpa Airport&
Normal take-off\\\hline
Last ACARS transmission & 7 Mar. 2014 17:07Z &
Everything normal, on path to Beijing\\\hline
Loss of secondary radar & 7 Mar. 2014 17:21:13Z &
Onboard transponder goes inactive; last civilian radar\\\hline
7 Mar. 2014 18:22:12Z &
Last Malaysian military radar contact &
MH370 shown tracking NW through Malacca Straits\\\hline
7 Mar. 2014 18:25Z-18:28Z &
Satellite Data Unit (SDU) log-on sequence &
If flying level, MH370 still tracking NW\\\hline
7 Mar. 2014 18:39-18:41Z &
Unanswered ground-to-air phone call &
If flying level, MH370 tracking southwards\\\hline
7 Mar. 2014 19:41Z to 8 Mar. 2014 00:11Z&
SATCOM metadata approx. hourly from MH370 & \cite{Ashton2014,Davey2016} suggest flight into Southern Indian Ocean\\\hline
8 Mar. 2014 00:19:29Z-00:19:37Z &
Partial SDU log-on sequence &
Implies rapid descent\\
\hline
\end{tabular}
\end{table*}

\section{Review of SATCOM Model}\label{sec:SATCOM_Model}
\begin{figure}[t]
\centering
\includegraphics[width=0.8\linewidth]{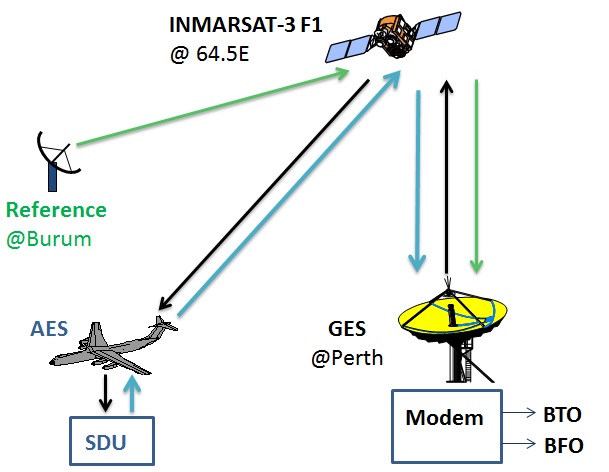}
\caption{System model of the satellite communication system \cite{Davey2016}.}\label{fig:SatSysMod}
\end{figure}

%
The accident aircraft was fitted with a SATCOM
terminal that used the Inmarsat Classic Aero system
\cite{Ashton2014}, which uses geosynchronous satellites to relay 
messages
between aircraft and ground stations. During the flight, messages were passed 
between the aircraft and a ground receiving
station located in Perth, Australia, via the Inmarsat-3F1 satellite.
Figure~\ref{fig:SatSysMod} illustrates the SATCOM
system in use during the flight. The aircraft is referred
to as the Aircraft Earth Station (AES) and the ground receiving unit
is referred to as the Ground Earth Station (GES). Inmarsat-3F1 is a
satellite in geosynchronous orbit at 64.5$^\circ$ East longitude 
and it was used exclusively for the duration of the flight.

An AES is equipped with an SDU comprising a satellite 
modem with auxiliary hardware and software. Transmission of data over the
satellite is via bursts which are scheduled to arrive at the GES at a 
specified time and a given frequency. As explained in 
\cite{Davey2016,Ashton2014},
communications from multiple users are coordinated by the allocation of 
different time and frequency slots to each user. This is done without knowledge
of individual AES locations or precise knowledge of the satellite location. 
Therefore, messages from a given AES might not
arrive at the GES at exactly the expected time, and generally would
arrive slightly later. The 
difference between the expected time of arrival (based on a nominal assumed 
position for the satellite and the AES) and the actual time of arrival
is referred to as the Burst Timing Offset (BTO). The BTO is a measure 
of how 
far the aircraft is from the sub-satellite position\footnote{The sub-satellite
position is the point on the earth directly below the satellite.}.

The relative velocity between the satellite and the AES, as well as between the
satellite and the GES, leads to 
a Doppler frequency offset on the signals received at the 
GES. Coupled with small frequency offsets inherent in the reference frequency
oscillators in the AES, satellite and GES, this results in a net difference 
between the expected and actual frequency of the signal
presented to the modem in the GES for a given user. Frequency compensations 
applied onboard 
the aircraft (aircraft induced Doppler pre-compensation) and at the ground 
station (Enhanced Automatic Frequency Correction, which utilizes the reference
signal transmitted from a reference station in Burum, Netherlands), \cite{Davey2016,Ashton2014} 
serve to reduce the possible difference between the expected and actual 
frequency of the messages received from the aircraft. The residual
difference between the expected frequency of each communications
burst and the actual received frequency is referred to as the BFO.
\subsection{Review of BFO Statistics}\label{sec:BFO_stats}
\begin{figure}[t]
\centering
\includegraphics[width=0.95\linewidth]{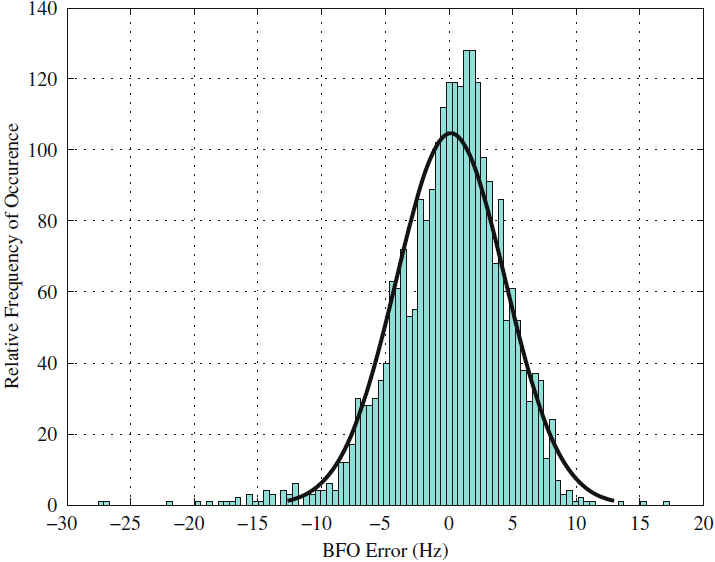}
\caption{Histogram of BFO errors for 20 flights of 9M-MRO prior to 
MH370 (reproduced from Fig.~5.5 of \cite{Davey2016}). The mean and standard deviation for the distribution are 0.18~Hz and 4.3~Hz, respectively \cite{Davey2016}.}\label{fig:BFO_hist}
\end{figure}

Based on 20 previous flights of 9M-MRO in the week leading up to the accident
flight (see \cite{Davey2016} for further details), a histogram was produced 
for the difference between the predicted BFO (based on known details of the 
plane and the satellite's position and velocity) and the measured BFO 
(based on Inmarsat ground station logs). This difference (i.e. predicted minus 
measured) is
referred to as the BFO error. The histogram of the BFO error is shown in 
Fig.~\ref{fig:BFO_hist}, along with a Gaussian distribution fit line. It can be 
seen that the distribution is somewhat
Gaussian. The standard deviation of the BFO error was found in 
\cite{Davey2016} to be 4.3~Hz. Whilst it is reasonable to apply 
bounds on the possible BFO error based on $\pm 3$ standard deviations as was 
done for the approximate analysis described in \cite{ATSB2016}, for the 
purpose of the descent analysis presented later in 
Sec.~\ref{sec:Descent}, it is assumed the BFO error is strictly bounded
on the larger interval $\left[-28,+18\right]$ Hz, which corresponds to the 
bounds of all 2501 observed valid\footnote{One outlier was removed as 
explained in Sec.~\ref{sec:OCXO_Detail}.} in-flight BFO error values available from the preceding 20 flights of 9M-MRO.

\section{Effects of Aircraft Position and Velocity on the Burst Frequency Offset}\label{sec:BFO_Detail}
\subsection{Mathematical Description of the BFO}\label{sec:BFO_maths}
In \cite{Davey2016}, the BFO is defined mathematically at time step $k$ as the 
sum of a 
noiseless component $h_k^{\mathsf{BFO}}$ and a scalar $w_k^{\mathsf{BFO}}$ that 
represents the BFO noise. The noiseless component of the BFO is defined in
\cite{Davey2016} as the sum:
\begin{eqnarray}
	\nonumber
	h_k^{\mathsf{BFO}}\left(\mathbf{x}_k,\mathbf{s}_k\right) &=&
	\Delta F_k^{\mathsf{up}}\left(\mathbf{x}_k,
	\mathbf{s}_k\right) + \Delta F_k^{\mathsf{down}}\left(\mathbf{s}_k\right)\\
	\nonumber
	&&+ \delta f_k^{\mathsf{comp}}\left(\mathbf{x}_k\right)
	+ \delta f_k^{\mathsf{sat}}\left(\mathbf{s}_k\right)\\
	&&+ \delta f_k^{\mathsf{AFC}}\left(\mathbf{s}_k\right)
	+ \delta f_k^{\mathsf{bias}}\left(\mathbf{x}_k,\mathbf{s}_k\right)\mbox{,}
	\label{eq:BFO_full}
\end{eqnarray}
where
\begin{itemize}
	\item $\mathbf{x}_k$ denotes the state vector of the aircraft;
	\item $\mathbf{s}_k$ denotes the state vector of the satellite;
	\item $\Delta F_k^{\mathsf{up}}\left(\mathbf{x}_k,
		  \mathbf{s}_k\right)$ is the uplink (aircraft to satellite) Doppler 
		  shift;
	\item $\Delta F_k^{\mathsf{down}}\left(\mathbf{s}_k\right)$ is the downlink 
	      (satellite to ground station) Doppler shift;
	\item $\delta f_k^{\mathsf{comp}}\left(\mathbf{x}_k\right)$ is the 
		  frequency compensation applied by the aircraft;
	\item $\delta f_k^{\mathsf{sat}}\left(\mathbf{s}_k\right)$ is the variation
	      in satellite translation frequency;
	\item $\delta f_k^{\mathsf{AFC}}\left(\mathbf{s}_k\right)$ is the frequency
	      compensation applied by the ground station receive chain;
	\item $\delta f_k^{\mathsf{bias}}\left(\mathbf{x}_k,\mathbf{s}_k\right)$ is
	      a slowly varying bias due to errors in the aircraft and satellite 
	      oscillators and processing in the SDU.
\end{itemize}

By treating the bias $\delta
f_k^{\mathsf{bias}}\left(\mathbf{x}_k,\mathbf{s}_k\right)$ as a constant 
determined at the source tarmac for any particular flight, as was done in 
\cite{Davey2016} for MH370, any small time-varying component of the bias during 
a particular flight
can be considered as part of the BFO noise (indeed this was done when compiling
the results used to obtain the BFO error histogram shown in
Fig.~\ref{fig:BFO_hist}). Details regarding the terms 
$\delta f_k^{\mathsf{sat}}\left(\mathbf{s}_k\right)$ and $\delta 
f_k^{\mathsf{AFC}}\left(\mathbf{s}_k\right)$ are provided in \cite{Ashton2014}.
Tabulated values of the sum of these two terms were provided by Inmarsat to the 
MH370 Flight Path Reconstruction 
Group to use in estimating the likely trajectory flown. These
two terms depend on the satellite state $\mathbf{s}_k$ only, and not on the
aircraft state $\mathbf{x}_k$. Moreover, the downlink Doppler $\Delta 
F_k^{\mathsf{down}}\left(\mathbf{s}_k\right)$ does not depend on the location or
velocity of the aircraft, and can be calculated given the
known frequency of the downlink and the known satellite state at 
any given time.

Equation~(\ref{eq:BFO_full}) can be then be simplified to:
\begin{eqnarray}
	\nonumber
	h_k^{\mathsf{BFO}}\left(\mathbf{x}_k,\mathbf{s}_k\right) &=&
	\Delta F_k^{\mathsf{up}}\left(\mathbf{x}_k,\mathbf{s}_k\right)
	+ \delta f_k^{\mathsf{comp}}\left(\mathbf{x}_k\right)
	+ \delta f_k^{\mathsf{det}}\left(\mathbf{s}_k\right)\mbox{,}\\
	&&\label{eq:BFO_simple}
\end{eqnarray}
where $\delta f_k^{\mathsf{det}}\left(\mathbf{s}_k\right)$ is effectively a known deterministic value 
for any time step $k$. The other terms in (\ref{eq:BFO_simple}) couple the 
aircraft state $\mathbf{x}_k$ by way of the aircraft position and velocity to
the BFO as per the following equations adapted from
\cite{Davey2016}\footnote{Note that the sign convention used in \cite{Davey2016} is opposite
to that used in (\ref{eq:UplinkDopp})}:
\begin{eqnarray}
	\Delta F_k^{\mathsf{up}}\left(\mathbf{x}_k,\mathbf{s}_k\right) &=&
	\frac{F^{\mathsf{up}}}{c}
	\frac{\left(\mathbf{v}_{\mathsf{s}} -
	\mathbf{v}_{\mathsf{x}}\right)^\mathsf{T}
	\left(\mathbf{p}_{\mathsf{x}} -
		\mathbf{p}_{\mathsf{s}}\right)}
	{\left|\mathbf{p}_{\mathsf{x}} -
			\mathbf{p}_{\mathsf{s}}\right|}\mbox{,}\label{eq:UplinkDopp}\\
	\delta f_k^{\mathsf{comp}}\left(\mathbf{x}_k\right) &=&
	\frac{F^{\mathsf{up}}}{c}
	\frac{\left(\hat{\mathbf{v}}_{\mathsf{x}}\right)^\mathsf{T}
	\left(\hat{\mathbf{p}}_{\mathsf{x}} -
	\hat{\mathbf{p}}_{\mathsf{s}}\right)}
	{\left|\hat{\mathbf{p}}_{\mathsf{x}} -
		\hat{\mathbf{p}}_{\mathsf{s}}\right|}\mbox{,}\label{eq:AESComp}
\end{eqnarray}
where the dependence on the time step $k$ on the right hand side of the 
equations has been removed for simplicity of notation. In equations 
(\ref{eq:UplinkDopp}) and (\ref{eq:AESComp}), $\left|\cdot\right|$ is the
three dimensional Cartesian distance, and:
\begin{itemize}
	\item $F^{\mathsf{up}}$ is the uplink carrier frequency;
	\item $c$ is the speed of light;
	\item $\mathbf{v}_\mathsf{s}$ is the velocity vector of the satellite;
	\item $\mathbf{v}_\mathsf{x}$ is the velocity vector of the plane;
	\item $\mathbf{p}_\mathsf{s}$ is the position vector of the satellite;
	\item $\mathbf{p}_\mathsf{x}$ is the position vector of the plane;
	\item $\hat{\mathbf{v}}_\mathsf{x}$ is the SDU's estimate of the plane's 
	velocity vector, which is obtained using the plane's track angle and ground 
	speed, whilst assuming the vertical speed is zero;
	\item $\hat{\mathbf{p}}_\mathsf{s}$ is the SDU's estimate of the position 
	vector of the satellite, which assumes the satellite is at its nominal 
	orbital slot of 0 degrees North and 64.5 degrees East;
	\item $\hat{\mathbf{p}}_\mathsf{x}$ is the SDU's estimate of the position 
	vector of the plane, which is obtained using the plane's latitude and 
	longitude, whilst assuming the plane is at sea level.
\end{itemize}
\subsection{Effect of uncompensated vertical velocity}\label{sec:DirectDopp}
\begin{figure}[t]
\centering
\includegraphics[width=0.95\linewidth]{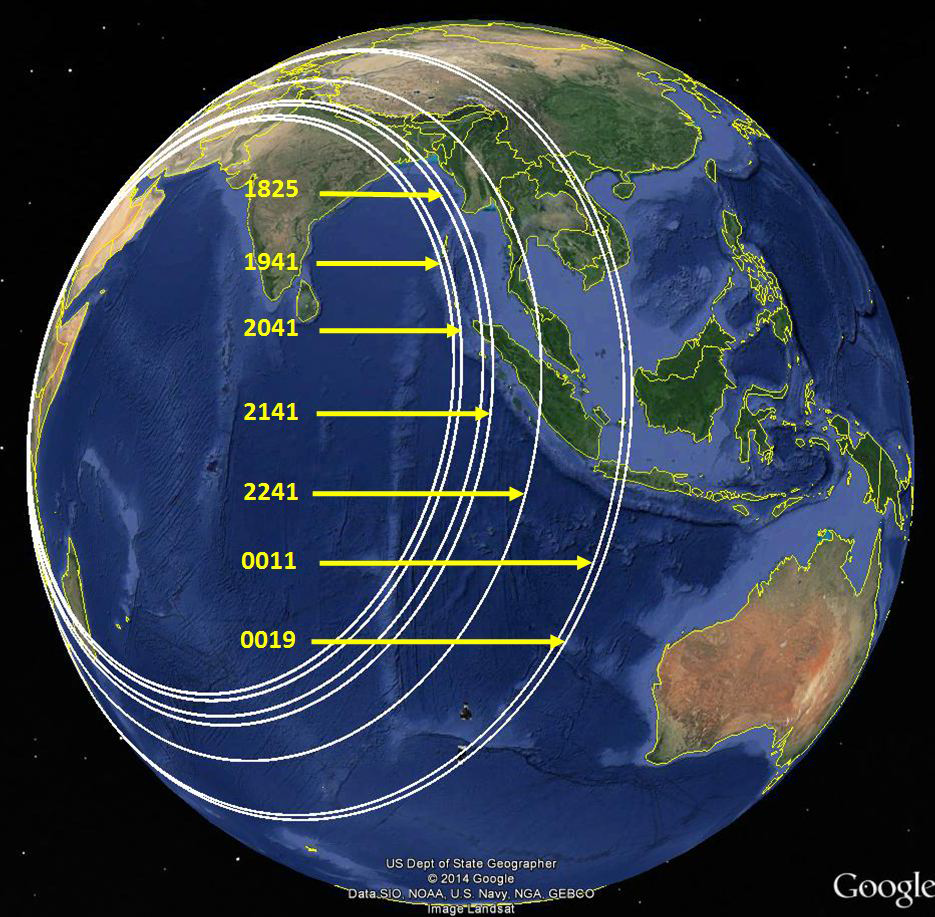}
\caption{BTO rings during the MH370 flight \cite{ATSB2014}.
Note that the latitude
of the arrows shown for each ring is arbitrary.}\label{fig:ATSB_Rings}
\end{figure}

The vertical speed of the plane is not used in the SDU Doppler compensation. As 
such, there is a direct contribution of Doppler due to the proportion of the 
vertical velocity vector projected onto the radial direction from 
the aircraft to the satellite. It is straightforward to understand that if the 
plane was directly below the satellite, the vertical velocity 
vector would be 
fully towards or away from the satellite if the plane was climbing 
or descending.
The direct contribution of the Doppler to the BFO in that case would be 
governed by the following standard Doppler equation.
\begin{eqnarray}
	\Delta F_k^{\mathsf{up}}\left(\mathbf{x}_k,\mathbf{s}_k\right) &=&
	\frac{v_z \cdot F^{\mathsf{up}}}{c}
	\mbox{,}\label{eq:DirectDopp}
\end{eqnarray}
where $F^{\mathsf{up}}$ and $c$ are as previously defined, and $v_z$ is the
vertical speed of the plane. Substituting an 
uplink frequency of 1646.6525~MHz (the uplink frequency stated in 
\cite{Ashton2014}) and a vertical velocity of 100 feet per minute (fpm), 
equivalent to 0.508 meters per second, equation (\ref{eq:DirectDopp}) implies
that the predicted BFO would increase by 2.8~Hz per 100~fpm of climb rate or
decrease by 2.8~Hz per 100~fpm of descent rate if 
the plane were directly below the satellite. This is the maximum possible 
contribution of the plane's climb or descent rate to the BFO. In the more 
general case, 
equation (\ref{eq:DirectDopp}) is moderated by the sine of the elevation angle  
$\theta$ from the aircraft to the satellite. This is expressed as
\begin{eqnarray}
	\Delta F_k^{\mathsf{up}}\left(\mathbf{x}_k,\mathbf{s}_k\right) &=&
	\frac{v_z \cdot F^{\mathsf{up}}\sin(\theta)}{c}
	\mbox{.}\label{eq:ProjectedDopp}
\end{eqnarray}
As such, at 00:19Z (the time at which the plane crosses the $7^{th}$ 
arc provided by Fig.~\ref{fig:ATSB_Rings})\footnote{The arcs 
referred to in this paper and other MH370 literature are segments of the rings 
in Fig.~\ref{fig:ATSB_Rings}.}, where the elevation 
to the satellite 
is $38.8$~degrees, the contribution to the BFO of climb or descent rate is 
reduced to approximately $+1.7$~Hz or $-1.7$~Hz per 100~fpm respectively.
\subsection{Effect of track angle on BFO towards end-of-flight}\label{sec:TrackAngle}
\begin{figure}[t]
\centering
\includegraphics[width=0.9\linewidth]{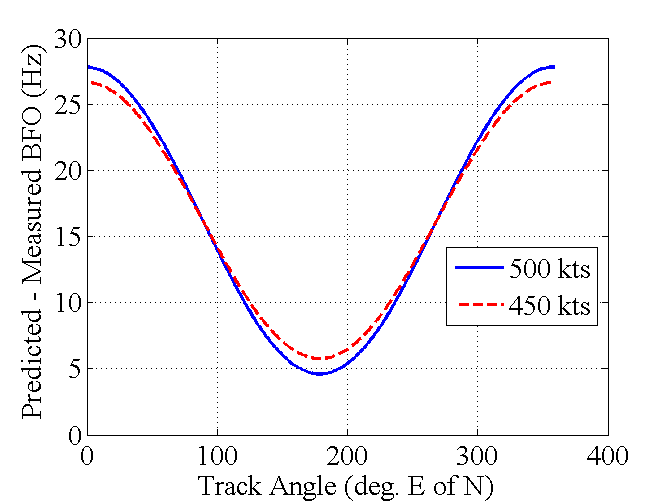}
\caption{BFO errors as a function of track angle at 
00:11Z.}\label{fig:TrackAngles0011}
\end{figure}

Another factor that needs to be considered in interpreting the
BFOs at 00:19Z is the track angle of the aircraft.
In \cite{Davey2016}, Fig.~5.6, a set of curves were shown illustrating
the relationship of the BFO error and the aircraft track angle at 18:39Z.
That was done to illustrate that under
the assumption of level flight during the unanswered telephone call period of
18:39Z-18:41Z, MH370 would have been tracking in a southerly direction. The same
model can be used to examine the relationship
between the BFO error and the track angle
at 00:11Z, when MH370 crossed the 6th arc, just 8 minutes prior to 
the last messages were received from MH370.\footnote{Note that
the 6th arc crossing is considered here to isolate effects of ground track
angle variation as opposed to vertical velocity, since the measured BFO at
the 6th arc is still consistent with level flight.}
To generate the curve of BFO error vs. track angle, it is 
assumed that
the aircraft crossed this arc at 38.67S, 85.11E. A previous sensitivity analysis
has revealed the curve is relatively insensitive to the actual 
crossing point. Also, calculations revealed that (as shown in Fig.~5.6 of
\cite{Davey2016} at 18:40Z) the track angle of the aircraft influences the
BFO to a greater degree for a faster assumed ground speed. To assess the
maximum possible effect of track angle on the BFO at 00:19Z, a 
ground speed
of 500~kts is assumed, and the BFO error vs. track angle is then as 
shown in Fig.~\ref{fig:TrackAngles0011}.
Another curve assuming 450~kts ground speed is shown for 
comparative purposes. It is noted that the peak-to-peak variation of the BFO
difference is similar for both considered ground-speeds.
\subsection{BFO Trend During the MH370 Flight}\label{sec:BFO_Trend}
\begin{figure}[t]
\centering
\includegraphics[width=0.95\linewidth]{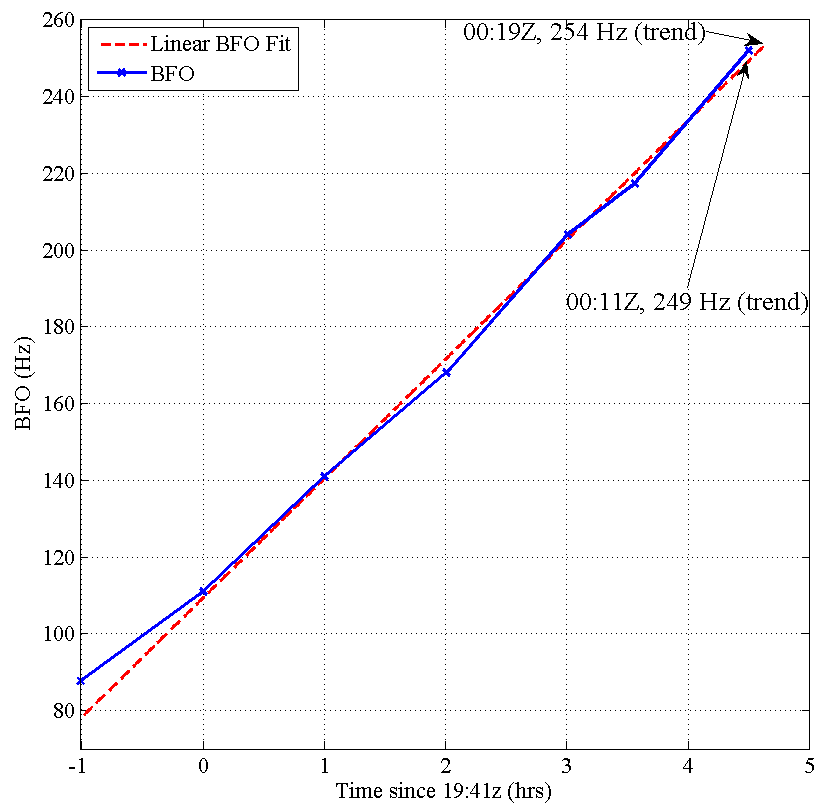}
\caption{Measured BFOs during the MH370 flight with trend-line extrapolated back to 18:40Z and forward to 00:19Z.}\label{fig:BFO_Trend}
\end{figure}

The measured BFOs from 19:41Z to 00:11Z are shown in Fig.~\ref{fig:BFO_Trend}
together with a line-of-best-fit. This line is extended forward 
to the 
time of the 00:19Z log-on resulting in an expected BFO of roughly 
254~Hz\footnote{Note the BFOs observed at 00:19Z were much lower than the 
expected value.}. With reference to the track angle curves 
presented in Fig.~\ref{fig:TrackAngles0011}, the 00:11Z BFO error 
value for the southern-most track is roughly 6~Hz, meaning the measured BFO of 
252~Hz was 6~Hz lower than the lowest value 
it could have been (assuming roughly level flight at approximately 450~kts).
If we assume the same error value at 00:19Z, and use the extrapolated BFO,
assuming a south track with level flight and similar ground speed, the expected 
BFO value is 260~Hz.

The linear trend-line in Fig.~\ref{fig:BFO_Trend} is also extended back by 
1~hour to the time of the first unanswered satellite telephone call to MH370. 
The mean of the measured BFOs during that call attempt is also 
shown. Since the
mean of the measured BFOs from the 18:39-18:41Z call attempt are in 
broad agreement with 
the linear trend observed in the BFOs from 19:41Z to 00:11Z (for which the 
BTOs themselves were consistent with straight and level flight 
\cite{Davey2016}), this supports the finding in \cite{Davey2016} that there 
were most-likely no major turns after the unanswered call attempt (see 
(\cite{Davey2016}, Fig.~10.5)).
\section{Effect of SDU Startup on the BFO}\label{sec:OCXO_Detail}
The BFOs logged for 
messages from the plane received at the Perth GES at 00:19:29Z and 00:19:37Z 8 
March 2014 were much lower than the expected value of
260~Hz. In the absence of any other factors affecting
the BFOs at these times, this suggests a large uncompensated negative vertical
velocity component in the BFOs. In determining how to interpret this data, it is
important therefore to carefully consider any known other factors that could 
effect these BFOs. One potential factor is an SDU oscillator
startup transient. This section analyses the range of
expected effect on the BFO caused by a period of power outage followed by a 
restart 
of the SDU, as was a possibility between 00:11Z and 00:19Z on 8 March 2014 
(see \cite{ATSB2014} for further details).
\subsection{Detailed Analysis of SDU Startup Effects on the BFO}
\begin{figure}[t]
\centering
\includegraphics[width=0.95\linewidth]{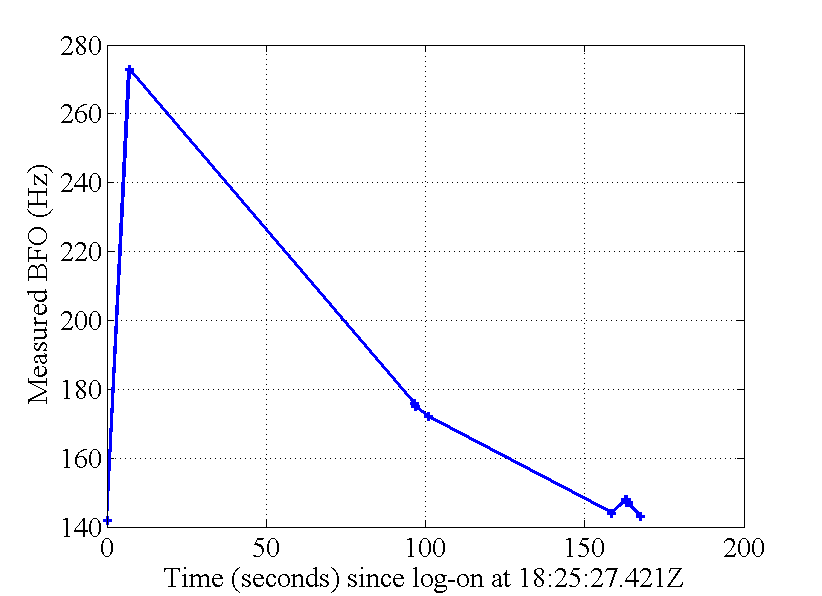}
\caption{Measured BFOs following the 18:25:27Z log-on message from 
MH370.}\label{fig:1825_logon_bfos}
\end{figure}

At 18:25:27Z 7 March 2014, the Inmarsat GES in Perth received a SATCOM log-on 
request from 
9M-MRO. A series of messages were exchanged in the following few minutes as 
part of a standard log-on sequence. The BFOs over those minutes displayed 
somewhat unusual behavior in that (barring the first BFO) the BFO appeared to
be exhibiting a transient settling behavior. This is shown in 
Fig.~\ref{fig:1825_logon_bfos}. It was noted in \cite{Ashton2014} that the 
spike in BFO observed at this time\footnote{When viewed on a time-scale of 
hours as shown in Fig.~9 of \cite{Ashton2014}, this behavior looks like a 
spike in BFO between 18:25Z and 18:28Z.} was not fully understood and whilst 
originally attributed to a possible aircraft maneuver it could also be related 
to actual frequency changes occurring during the logon sequence. A subsequent 
study \cite{SWG2016} by the SATCOM sub-group of the MH370 Flight Path
Reconstruction Group revealed this was most 
likely due to power-on frequency drift and subsequent stabilization of the 
oven-controlled crystal oscillator (OCXO) in the SDU. In
\cite{SWG2016}, a number of different SDUs were tested to
investigate the effects of power outages on the BFO during SDU power 
on, triggering a SATCOM logon such as that which occurred for MH370 at 
18:25:27Z. It was found that whilst the frequency settling behavior was 
different for each individual SDU, any given SDU displayed repeatable behavior 
for a fixed outage duration, and similar settling characteristics for outages
of different duration.

Based on sequences of Inmarsat BFO logs for 9M-MRO concerning four log-ons 
after periods of SATCOM outage in the week leading up to the accident flight, 
it was found (in \cite{SWG2016}) that the SDU in 9M-MRO resulted in a BFO that
was initially too high at the time
of log-on, followed by a simple decay over the next few minutes to a 
steady-state value. Regarding the log-on event at 18:25:27Z for MH370, it was 
noted that there was a non-zero bit error rate (BER) associated with the 
log-on 
request at that time. The associated received signal level and carrier-to-noise
density ratio ($C/N_0$) were also unusually low. As such, the first BFO was 
deemed 
untrustworthy.\footnote{An alternative explanation for this outlier BFO
could be the unknown conditions of the aircraft between 17:22Z and 18:25Z, for instance the temperature of the SDU might have been much lower than the expected ambient temperature.}
Note that this is not the only time such a behavior was 
observed for the SDU in 9M-MRO. In the analysis of BFOs presented in 
\cite{Davey2016}, there was an outlier in the BFO (with an error of -170 Hz)
found to have occurred for a SATCOM message associated with 9M-MRO in
one of the 20 previous flights of 9M-MRO. This was discarded as an 
outlier and not presented in \cite{Davey2016}. It too had a non-zero BER and
a low $C/N_0$ (37.6~dBHz compared to surrounding values between 41.5 and 
42~dBHz).
With the removal of the untrustworthy BFO logged for the 
first message in the sequence shown in Fig.~\ref{fig:1825_logon_bfos}, the 
18:25Z log-on from MH370 was also determined to follow the simple decay trend 
observed in other instances.

\begin{figure}[t]
\centering
\includegraphics[width=0.95\linewidth]{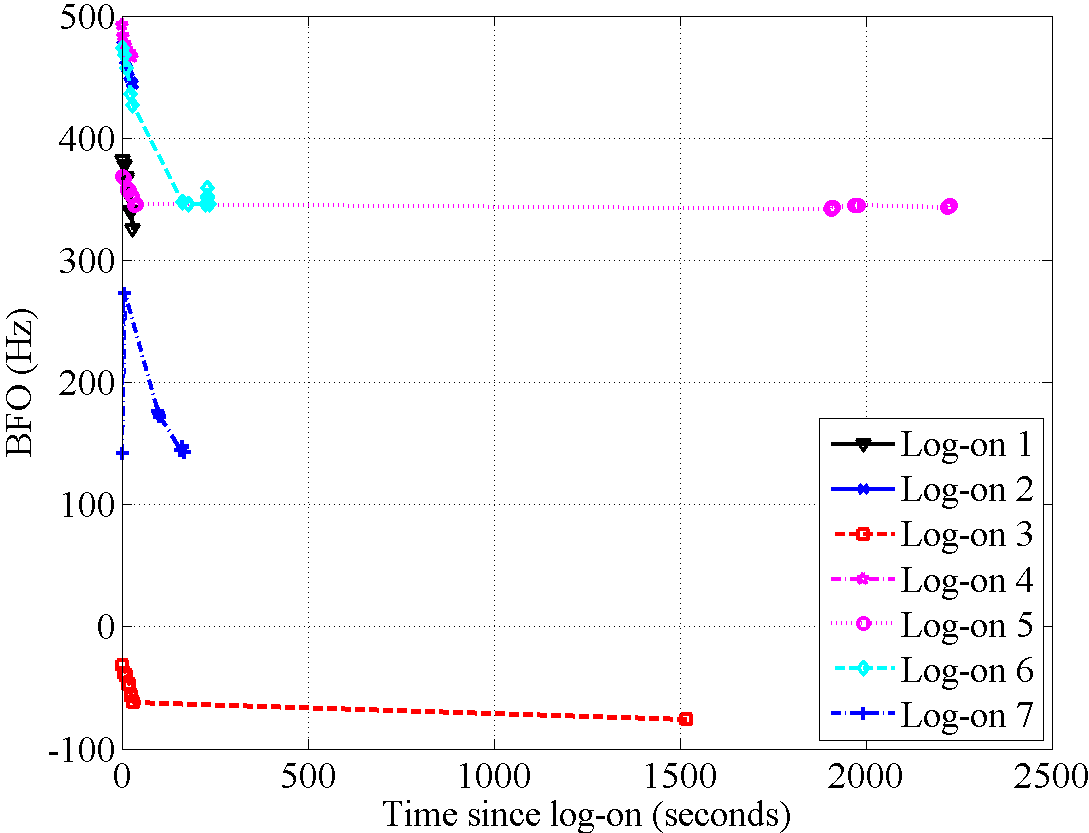}
\caption{Measured BFOs for 7 log-ons of 9M-MRO.}\label{fig:raw_logon_BFOs}
\end{figure}

It was noted by DST Group that if the simple BFO decay trend established after
SDU power-up \cite{SWG2016} 
also occurred during the MH370 SATCOM log-on event beginning at 00:19:29Z, then 
it would allow bounds to be established on the BFO that
would have been expected for MH370 if the plane was flying level at that 
time. This in turn would allow the determination of bounds on the possible 
descent rates of MH370 during that final log-on event. In order to build
confidence that the simple BFO decay trend would hold for the 00:19Z log-on
event, DST Group reviewed additional Inmarsat logs for 9M-MRO, corresponding to
the period 22 February to 28 February 2014. Two additional cases were 
identified in which 9M-MRO logged back onto an Inmarsat satellite after a 
sustained period of SATCOM outage. The sequence of BFOs observed in these two 
cases, along with the five already considered by the SATCOM working group in 
\cite{SWG2016}, are shown in Fig.~\ref{fig:raw_logon_BFOs}, and details
about each of the log-ons are given in Table~\ref{tab:raw_logon_BFOs}.

\begin{table*}[!t]
\renewcommand{\arraystretch}{1.3}
\caption{Details of Log-on Sequences Used for Analysis}
\label{tab:raw_logon_BFOs}
\centering
\begin{tabular}{|c||c|c|c|}
\hline
Identifier & Date and Timestamp of Log-on & Duration of Preceding Power Outage & Any Other Comments\\
\hline\hline
Log-on 1 & 23rd Feb. 23:57Z & Between 381 and 442 minutes &
After scheduled A1 maintenance check, some non-zero BERs\\\hline
Log-on 2 & 26th Feb. 14:11Z& Between 295 and 354 minutes&\\\hline
Log-on 3 & 5th Mar. 03:06Z& Between 35 and 95 minutes&\\\hline
Log-on 4 & 6th Mar. 13:29Z& Between 43 and 103 minutes&\\\hline
Log-on 5 & 6th Mar. 15:02Z& Between 35 and 92 minutes&\\\hline
Log-on 6 & 7th Mar. 12:50Z& Between 228 and 288 minutes& Some non-zero 
BERs\\\hline
Log-on 7 & 7th Mar. 18:25Z& Between 20 and 78 minutes& First point 
untrustworthy\\
\hline
\end{tabular}
\end{table*}
The periods of SATCOM outages followed by log-on events were identified from
Inmarsat-provided ground-station logs by identifying sequences of three or more 
unsuccessful log-on interrogations to 9M-MRO (suggesting the SDU was likely 
powered off) followed by a log-on to the satellite system initiated from 
9M-MRO at some later time. The exact duration of the power-off period for the
SDU was unable to be determined from the logs, however the timestamps 
associated with the unsuccessful log-on interrogations and subsequent 9M-MRO
initiated log-ons were used to determine bounds on the outage time as shown in
the $3^{rd}$ column of Table~\ref{tab:raw_logon_BFOs}.\footnote{Note 
that the bounds shown are derived under the assumption that the 
SATCOM outage was exclusively due to a 
power outage. It is also possible that the first part of some outages was due 
to a blocked line of sight to the Inmarsat satellites.}
Additional comments 
about each log-on sequence are given in the final column. Whilst some of the 
BERs
for log-ons 1 and 6 were non-zero, the BFOs did not appear to be outliers, and 
were therefore considered valid.
The uncertainty window in the length of the 
outage durations shown is roughly 60 minutes;
this relates to timers in the Inmarsat satellite ground stations that result
in a log-on interrogation being sent to the plane after roughly 60 minutes of 
SATCOM inactivity. Indeed, it is this same timing responsible for the roughly 
60 minute separation between the four BTO arcs from 19:41Z to 22:41Z (e.g. 
\cite{ATSB2014,Ashton2014}). It is also of note that based on other evidence 
related to MH370 (e.g. secondary radar transponder loss of signal at 17:22Z) 
that SDU power was probably first lost at 17:22Z, in which case the SDU power 
outage duration for log-on 7 would be approximately 63 minutes.

\begin{figure}[t]
\centering
\includegraphics[width=0.95\linewidth]{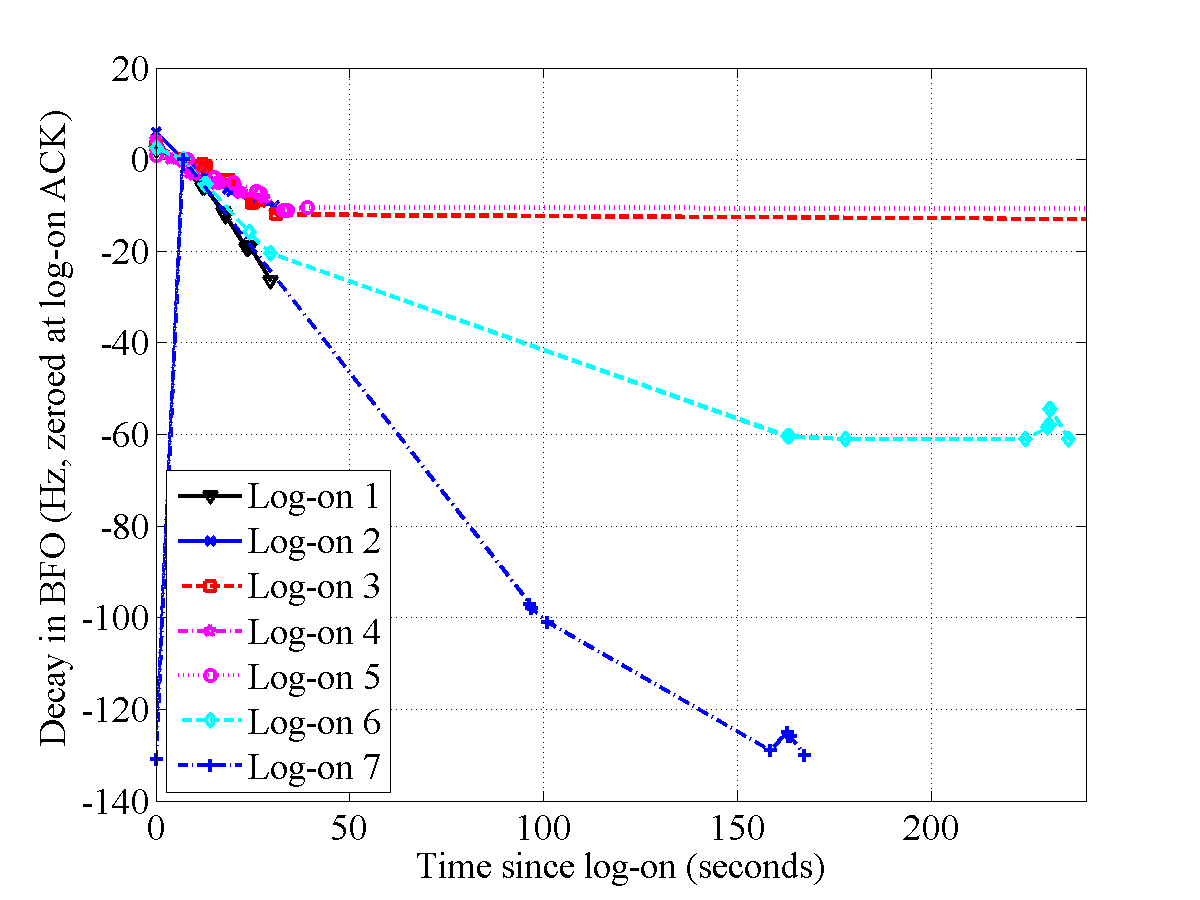}
\caption{Decay in measured BFOs for 7 log-ons of
9M-MRO. The decay rates for log-ons 1 to 6 have been adjusted
to account for the different AES Doppler compensation mode
used during these log-ons.}\label{fig:logon_bfo_decays}
\end{figure}

For log-ons 3 and 5, there were some points available about half an hour
after the log-on in each case (noting the plane was on the tarmac for the 
whole time) that demonstrate the BFO appears to have fully settled to a 
steady-state value. Log-ons 1 to
5 only lasted for less than 1 minute after the log-on request message. The 
results for log-ons 6 and 7, which had log-on sequences lasting a few minutes 
suggest a settling behavior of approximately 3 minutes duration. For all
observed log-on sequences, there seems to be a simple decay of BFO over a few
minutes after SDU startup and log-on request (see also 
Fig.~\ref{fig:logon_bfo_decays}). It is therefore possible a similar
decay was occurring during the SDU log-on sequence beginning at 00:19:29Z.

In order to establish what the likely range of steady-state equivalent BFOs 
would have been at 00:19Z, it is useful to re-plot the curves from 
Fig.~\ref{fig:raw_logon_BFOs} translated vertically such that the log-on
acknowledgment has 0 Hz offset. This is done in 
Fig.~\ref{fig:logon_bfo_decays}, which has also been zoomed in the time-axis
for clarity.

For log-ons 1 to 6, the aircraft was on the ground at an airport. 
It has been 
established that the most common scenario for 
an on-ground-at-airport log-on after sustained power 
outage of the SDU is for
the AES to use a different self-Doppler compensation method than was in use for 
log-on 7 and the log-on at 00:19Z. This is due to a lack of navigational input 
data from the aircraft's inertial navigation unit in the few minutes after 
power-up, and is referred to as the closed-loop Doppler compensation
approach.\footnote{The use of closed-loop Doppler compensation in this 
circumstance was
established in consultation with members of the MH370 Flight Path Reconstruction
Group, including representatives from Thales and Inmarsat.}

The standard AES Doppler compensation mode when navigational data is 
available to the SDU is the mode used for log-on 7 and the log-on at 00:19Z. 
This approach was briefly described in Sec.~\ref{sec:BFO_maths}. It used the 
aircraft's own estimates of its position and horizontal velocity to predict
and correct the AES contributed Doppler shift. This is
referred to as open-loop Doppler compensation.

For log-ons 1 to 6, we work on the premise that closed-loop Doppler 
compensation was used.\footnote{Previous 
analysis had not considered the difference in Doppler compensation mode, yet 
still resulted in similar descent rates to those presented in this paper.}
In 
that approach, the frequency of broadcast P-channel transmissions received by 
9M-MRO 
were used to determine the AES Doppler contribution. In this approach, all of 
the difference between the received and expected P-channel frequency is 
attributed by the SDU to be caused by AES motion induced Doppler. As such, if 
the OCXO frequency is 
higher than its nominal settled value, due to incomplete temperature 
stabilization, the received frequency will appear lower than expected by the 
corresponding amount. When applying the AES Doppler compensation in 
this case, the transmit frequency is increased by the difference between the 
stable and pre-stabilized oscillator frequency (scaled to account for the 
difference between P-channel frequency and the R-channel transmit frequency). 
Moreover, the transmit 
frequency itself (prior to AES Doppler compensation) will already be too high 
by the same amount due to the incomplete temperature stabilization. The net
combined effect to the BFO decay observed due the OCXO settling will therefore
be doubled compared to the case where open-loop AES Doppler pre-compensation is 
used. Hence, the BFO decay results for log-ons 1 to 6 shown in 
Fig.~\ref{fig:logon_bfo_decays} have 
been halved for the analysis when compared to those inferred from the raw 
measured BFOs presented in Fig.~\ref{fig:raw_logon_BFOs}.

Recall that the two messages at 00:19:29Z and 00:19:37Z are a log-on and log-on
acknowledgment respectively. From the data used to plot 
Fig.~\ref{fig:logon_bfo_decays}, it can be determined
that the log-on acknowledgment BFOs are in the range of 
[0,6]~Hz lower than the log-on BFOs. Also
the maximum difference between the log-on BFO and the settled 
BFO value is in the range [17,136]~Hz. The only log-on that doesn't appear to 
be approaching a settled value in this range is log-on 1, for which it is 
anticipated that if more data points were available several minutes later, as 
was the case for log-ons 3, 5, 6 and 7, the settling behavior would be 
similar to that for log-ons 6 and 7, bearing in mind the outage duration for 
log-on 1 is probably at least 381 minutes, and the BFO already dropped 
substantially in the first 30 seconds after log-on.
\subsection{Summary of SDU Startup Effects on the BFO}\label{sec:SDU_StartupSum}
It is likely that the SDU in 9M-MRO lost power sometime after 00:11Z, and then 
regained power (presumably due to Auxiliary Power Unit (APU) startup) prior to 
00:19Z, leading to the SDU making the log-on attempt at 00:19:29Z. Previous 
such events for 9M-MRO have 
shown that this results in an initially too-high BFO for the log-on message, 
followed by a simple decay characteristic to reach a steady-state BFO after 
several minutes. In order to establish the relevance of the BFOs at 00:19:29Z 
and 00:19:37Z to
the motion (including descent rate) of MH370 at those times, this section has 
established bounds for the likely range of steady-state equivalent BFOs at 
those times. This was done by analyzing BFOs from seven previous SDU start-up 
events for a period of several minutes. The specific results obtained suggest that:
\begin{enumerate}
	\item The recorded BFO for the 00:19:29Z log-on was 
	between 17 and 136~Hz higher than it would have been if the OCXO in the SDU
	was in a steady state at that time;
	\item The recorded BFO for the 00:19:37Z log-on acknowledge
	message was between 17 and 130~Hz higher than it would have been if the OCXO
	in the SDU was in a steady state at that time.
\end{enumerate}
\section{Bounding the Descent Rates of MH370}\label{sec:Descent}
The results from Sections~\ref{sec:BFO_Detail} and \ref{sec:OCXO_Detail} can
be combined to provide bounds on the descent rate of MH370 implied
by the BFOs from the two last SATCOM messages for the flight. In this section, 
it is shown how this is done for 
two different possibilities that could explain the attempted SATCOM log-on
from 9M-MRO at 00:19Z. In \cite{ATSB2014} the most likely cause of this 
log-on was stated to be a power interruption resulting from insufficient fuel
and subsequent engine flameout. It could also have been due to a temporary 
software failure, a loss of systems providing critical input to the SDU, or a 
loss of the SATCOM link due to aircraft attitude being such that the 
line-of-sight to the satellite is blocked. If it was indeed a power 
interruption to the SDU caused by loss of fuel and subsequent reboot using the
APU, the SDU would be without 
power for about one
minute. In this case, the results of Sec.~\ref{sec:SDU_StartupSum} need to be 
considered when interpreting the last two BFOs.\footnote{Given the relatively 
short duration of the power-loss under Hypothesis 1, it is possible that the 
extent of BFO decay due to OCXO warm-up would be less than shown for log-ons 1 
to 7. This is essentially covered within Hypothesis 2, which considers a 
momentary SATCOM outage.} If on the 
other hand, the 
power loss was momentary (resulting in a reset of the SDU) or if the 
temporary SATCOM outage leading to the log-on request was due to one of the
other listed reasons, there would be no ``warm-up drift'' to consider, so the
results of Sec.~\ref{sec:OCXO_Detail} would not need to be applied. Both cases 
are considered separately in the following two sub-sections, and then combined 
overall bounds are presented.
\subsection{Hypothesis 1: SATCOM outage due to insufficient fuel}
In the event that the SDU log-on at 00:19:29Z was due to engine flame-out,
followed by a restart of the SDU using power from the APU, the SDU outage 
preceding
the log-on would have lasted about one minute. This would result in some 
cooling of the OCXO in the SDU.
The likely effect of the SDU startup and consequent `warm-up drift' of the OCXO
is summarized in Sec.~\ref{sec:SDU_StartupSum}.
Table~\ref{tab:last_2_bfos_Hyp1} presents the recorded BFOs and bounds on the 
adjusted BFOs to remove these effects of warm-up drift. In the last column of 
the table, the bounds are extended taking into account the BFO noise bounds of
$\left[-28,+18\right]$ Hz established in Sec.~\ref{sec:BFO_stats}.
\begin{table*}[!t]
\renewcommand{\arraystretch}{1.3}
\caption{The Last Two BFOs for MH370 Under Hypothesis 1}
\label{tab:last_2_bfos_Hyp1}
\centering
\begin{tabular}{|c||c|c|c|}
\hline
Timestamp & Recorded BFO (Hz) & BFO Range if Start-Up Drift Removed (Hz) & 
Extended Range Considering BFO Noise (Hz)\\
\hline\hline
00:19:29Z & 182 & $\left[46,165\right]$ & $\left[28,193\right]$\\
\hline
00:19:37Z & -2 & $\left[-132,-19\right]$ & $\left[-150,9\right]$\\
\hline
\end{tabular}
\end{table*}

As established in Sec.~\ref{sec:DirectDopp}, the recorded BFO would be roughly 
1.7~Hz lower for every 100~fpm of descent rate. As such, 
depending
on whether the plane was tracking South or North (minimum or maximum expected 
BFOs, respectively, see Sec.~\ref{sec:TrackAngle}), bounds on the descent rate
of MH370 at the times of transmission corresponding to the last 2 BFOs can be determined
by subtracting the values given in the rightmost column of 
Table~\ref{tab:last_2_bfos_Hyp1} from the expected BFO values for level flight
tracking South or North, dividing the result by 1.7~Hz and multiplying by 
100~fpm. The expected BFO for a south track is approximately 260~Hz, whilst for 
a north track it is close to 280~Hz. Using these numbers, it is straightforward
to obtain the bounds shown in Table~\ref{tab:Hyp1Bounds}. Note that the bounds
have been rounded to the nearest 100~fpm. Looking at all values in the table, it
can be concluded that irrespective of ground track angle and for assumed
ground speeds less than approximately 500~kts, under
Hypothesis 1, MH370 would have been descending at between 3,900 and 
14,800~fpm at 00:19:29Z, and just 8 seconds later at between 14,800 and 
25,300~fpm. These descent rates are consistent with
simulations of an uncontrolled phugoid descent reported in \cite{ATSB2016}.
\begin{table*}[!t]
\renewcommand{\arraystretch}{1.3}
\caption{MH370 Descent Rates at 00:19Z 8 March 2014 Under Hypothesis 1}
\label{tab:Hyp1Bounds}
\centering
\begin{tabular}{|c||c|c|c|c|}
\hline
Timestamp & Min. Desc. Rate, South Track & Min. Desc. Rate, 
North Track & Max. Desc. Rate, South Track & Max. Desc. Rate, 
North Track\\
\hline\hline
00:19:29Z & 3,900~fpm & 5,100~fpm & 13,600~fpm & 14,800~fpm\\
\hline
00:19:37Z & 14,800~fpm & 15,900~fpm & 24,100~fpm & 25,300~fpm\\
\hline
\end{tabular}
\end{table*}

\subsection{Hypothesis 2: SATCOM outage due to some other reason}
It is still of interest to determine bounds
on the descent rates under the alternate hypothesis that something else
not associated with a power outage led
to the SDU-initiated log-on event at 00:19Z. In this case, the warm-up drift
would not apply. Therefore, the recorded BFOs can be 
treated normally (though still subject to BFO noise). In this case, the BFO 
bounds are as set out in Table~\ref{tab:last_2_bfos_Hyp2}. Using these bounds, 
the lower and upper descent rates can be calculated again, 
resulting in the bounds presented in Table~\ref{tab:Hyp2Bounds}. It can be 
concluded from the table that the recorded 
BFOs indicate that at 00:19:29Z the plane was descending at between 2,900~fpm 
and 6,800~fpm. At the time of the last SATCOM message, it can be concluded 
under Hypothesis 2 that the descent rate would have been between 13,800~fpm and 
17,600 fpm.
\begin{table}[!t]
\renewcommand{\arraystretch}{1.3}
\caption{The Last Two BFOs for MH370 Under Hypothesis 2}
\label{tab:last_2_bfos_Hyp2}
\centering
\begin{tabular}{|c||c|c|}
\hline
Timestamp & Recorded BFO (Hz) & BFO Bounds With Noise (Hz)\\
\hline\hline
00:19:29Z & 182 & $\left[164,210\right]$\\
\hline
00:19:37Z & -2 & $\left[-20,26\right]$\\
\hline
\end{tabular}
\end{table}

\begin{table*}[!t]
\renewcommand{\arraystretch}{1.3}
\caption{MH370 Descent Rates at 00:19Z 8 March 2014 Under Hypothesis 2}
\label{tab:Hyp2Bounds}
\centering
\begin{tabular}{|c||c|c|c|c|}
\hline
Timestamp & Min. Desc. Rate, South Track & Min. Desc. Rate, 
North Track & Max. Desc. Rate, South Track & Max. Desc. Rate, 
North Track\\
\hline\hline
00:19:29Z & 2,900~fpm & 4,100~fpm & 5,600~fpm & 6,800~fpm\\
\hline
00:19:37Z & 13,800~fpm & 14,900~fpm & 16,500~fpm & 17,600~fpm\\
\hline
\end{tabular}
\end{table*}

\subsection{Summary of Descent Rate Bounds}
Combining the descent rate bounds for Hypotheses 1 and 2, outer bounds can
be established for the descent rate of MH370 at 00:19:29Z and 00:19:37Z. 
Specifically, regardless of ground track angle and for ground 
speeds less than approximately 500~kts, 
accounting for possible BFO noise, and regardless of whether the SATCOM outage
between 00:11Z and 00:19Z was due to an SDU power outage or another reason, the 
outer bounds on the possible descent rates at the times of the last two
SATCOM messages from MH370 are given in Table~\ref{tab:CombinedBounds}.
\subsection{Estimated Downwards Acceleration}
When interpreting the bounds presented in Table~\ref{tab:CombinedBounds}, the 
different conditions under which each bound was
derived need to be considered. For instance the lowest rate at 00:19:29Z was 
derived assuming no period of SDU outage and a southwards track, whereas the 
highest descent rate for 00:19:37Z was derived assuming a one minute outage of 
the SDU with the maximum BFO decay observed in the previous 7 SDU outage events 
for 9M-MRO. Hence, it would not be reasonable to say that the plane could have 
been descending at 2,900~fpm at 00:19:29Z and 25,300~fpm 8 seconds later, which 
would imply a downwards acceleration on the order of 2,800 fpm per second. 
Whilst it is not possible to determine a precise acceleration value, a rough
approximation of the average descent rate over those 8 seconds could for 
instance be taken using the mid-points of the bounds at each time, which would
result in an average downwards acceleration of 10,700~fpm in 8 seconds or 
around 1,300 fpm per second, which equates to around 6.7 ms$^{-2}$ or $0.68g$, 
where $g$ denotes Earth's gravitational constant, which is approximately 9.8 
ms$^{-2}$.
It is straightforward to see that other reasonable methods of estimating the 
downward acceleration (such as taking the difference between the two minimum 
descent rates or the two maximum descent rates at 00:19:29Z and 00:19:37Z) 
would yield similar acceleration values.
\begin{table}[!t]
\renewcommand{\arraystretch}{1.3}
\caption{Range of Possible MH370 Descent Rates at 00:19Z 8 March 2014}
\label{tab:CombinedBounds}
\centering
\begin{tabular}{|c||c|c|}
\hline
Timestamp & Min. Desc. Rate & Max. Desc. Rate\\
\hline\hline
00:19:29Z & 2,900 fpm & 14,800~fpm\\
\hline
00:19:37Z & 13,800 fpm & 25,300 fpm\\
\hline
\end{tabular}
\end{table}
\section{Conclusion}\label{sec:Conclusions}
This article has discussed in detail
all known factors that could have influenced the BFOs associated 
with the last two SATCOM messages received from MH370 at 00:19Z.
Lower and upper bounds on its descent rate at this time were then 
derived.\footnote{The
conclusions regarding descent rate have been made taking into account
BFO noise bounds, all reasonably feasible ground tracks and speeds, and possible
OCXO warm-up drift. It has been implicitly assumed that there were no otherwise
unknown factors that could have affected the last two BFOs.} The
downwards acceleration over the 8 second interval between these two messages
was found to be approximately $0.68g$. The derived bounds and approximate 
downwards 
acceleration rate are consistent with simulations of an uncontrolled descent 
near the $7^{th}$ arc as reported in \cite{ATSB2016}. This suggests that 9M-MRO 
should lie relatively close to the $7^{th}$ BTO arc.

At the time of submission of this paper, MH370 had not been located. 
The search was in indefinite suspension pending `credible new
evidence pointing to a specific location for the plane'. Further efforts in 
data 
analysis and in drift modeling of located MH370 debris have been conducted and 
reported (e.g. \cite{Griffin2016,Griffin2017,ATSB2016}). Note that the descent 
analysis from 
this paper was used in \cite{ATSB2016} when considering the width of any 
potential future search area.
Some recent news reports have suggested
a US underwater exploration firm named Ocean Infinity may soon be launching a
renewed search effort for MH370.
The reader should refer to the Australian Transport 
Safety Bureau website \cite{ATSBWeb} and the general media
for updates.
%
\section*{Acknowledgment}
The author would like to thank his Defence Science and Technology Group
colleagues Dr. Neil Gordon, Dr. Samuel Davey, Dr. Jason Williams, and Dr.
Mark Rutten who worked alongside him as part of the MH370 Flight Path
Reconstruction Group, as well as Mr. Balachander
Ramamurthy for his support to the satellite data analysis.
Special thanks goes to Dr. Gerald Bolding, both for his support to
the satellite data analysis and his extensive efforts in reviewing 
and improving this article. The author also thanks the many other members of 
the MH370 Flight Path Reconstruction Group, which comprised 
representatives from Inmarsat, Thales,
Boeing, US National Transportation Safety Board (NTSB), UK Air
Accidents Investigation Branch (AAIB), Honeywell, SED, Square Peg Communications Inc., and Panasonic. The group has been expertly led by the
Australian Transport Safety Bureau (ATSB).


%

%

\begin{IEEEbiography}[{\includegraphics[width=1in,height=1.25in,clip,keepaspectratio]{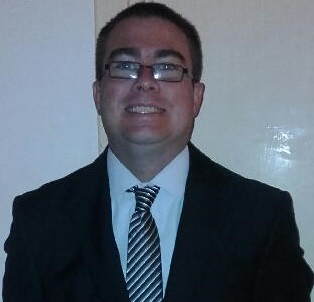}}]{Ian~D.~Holland}
received the Bachelor of Electronic and Communication
Engineering in 2000 and a Ph.D. in wireless telecommunications in 2005, both
from Curtin University of Technology, Western Australia. Since then he has
held positions in the University of Western Australia, the Institute for
Telecommunications Research at the University of South Australia, EMS Satcom
Pacific and Lockheed Martin Australia. Since January 2011, Ian has been
working as a Research Scientist in Protected Satellite Communications at the
Defence Science and Technology Group, Australia.
\end{IEEEbiography}

\ifCLASSOPTIONcaptionsoff
  \newpage
\fi


\end{document}